\documentclass[11pt,a4paper]{elsarticle}
\usepackage[margin=2cm]{geometry}
\usepackage{subfig}
\usepackage{amsmath,amsfonts,amssymb}
\usepackage{xcolor}

\begin{document}

\begin{frontmatter}
\title{Automatic detection of equiaxed dendrites using computer vision neural networks}

{\author{A.~Viardin \corref{cor}}}
\author{K.~N\"oth}
\author{M. Torabi~Rad}
\author{L.Sturz}
\address{ACCESS e.V., Intzestrasse 5, D-52072 Aachen, Germany.}

\cortext[cor]{Corresponding author}


\begin{abstract}
Equaixed dendrites are frequently encountered in solidification. They typically form in large numbers, which makes their detection, localization, and tracking practically impossible for a human eye. In this paper, we show how recent progress in the field of machine learning can be leveraged to tackle this problem and we present computer vision  neural network to automatically detect equiaxed dendrites. Our network is trained using phase-field simulation results, and proper data augmentation allows to perform the detection task in solidification conditions entirely different from those simulated for training. For example, here we show how they can successfully detect dendrites of various sizes in a microgravity solidification experiment. We discuss challenges in training such network along with our solutions for them, and compare the performance of neural network with traditional methods of shapes detection.
\end{abstract}

\end{frontmatter}

\section{Introduction}
 
Applying methods of artificial intelligence (AI), in particular "deep learning" for software-controlled 
automated identification, classification and characterization of various characteristics in materials, 
based on two-dimensional images coming from experimental results has shown a great interest in recent 
years \cite{holm2020}. In case of \emph{in situ} experimental pictures coming from video sequences, one can have several 
thousand of pictures to treat in order to extract physical features that are relevant. For this purpose, 
innovative AI tools must to be developed to overcome the limitations of classical methods of image processing.
For solification, several kind of microstructures can be observed during experiments, many of them are dendrites with
columnar or equiaxed morphology. The equiaxed dendrites are appearing in isothermal cooling conditions (they are 
growing in crystallographic directions), while columnar dendrites are growing mainly along a temperature gradient in 
an elongated form. 
The objective of this work is the automated detection and characterization of equiaxed
dendrites in a melt from time series using deep learning methods. 
Methods for detection an evaluation for microstructures using machine 
learning or deep learning techniques are already existing for solidification \cite{Liotti2018} and solid 
state transformations \cite{agbozo2020}. These works shows that neural networks for images 
detection is already working flawlessly. 
Deep learning model based on architecture like Faster R-CNN \cite{fasterrcnn}, Mask R-CNN 
\cite{maskrcnn} or Unet \cite{unet} have prooved their ability to detect objects 
with high accuracy. Nevertheless, the good results concerning objects detection/segmentation/instantiation are 
depending on the data used to train the network and can be very tedious to generate manually. Especially if one picture
contains hundred of objects to detect, annotations of these objects will be very time consuming. 
The objective of this article is to generate a huge amount of realistic microstructures 
to train deep learning models. This will be applied to experimental pictures obtained under reduced gravity with 
resting, focused, growing and later overlapping or connected equiaxed dendrites (experiment "MEDI-2" \cite{medi2}).
Recently Hsu \cite{hsu2020}, used generative adversarial network to generate 3D microstructures of solid oxide 
fuel cell electrodes. Here, we have used phase field simulation to generate pictures for training computer 
vision deep learning architectures in order to analyse experimental microstructures. Phase field models became 
an important tool for microstructure simulations in the field of solidification with or without 
fluid flow, solid state transformations or mechanics for example.

\section{Experimental results}

\begin{figure}[htbp]
  \centering
  \includegraphics[width=0.93\columnwidth]{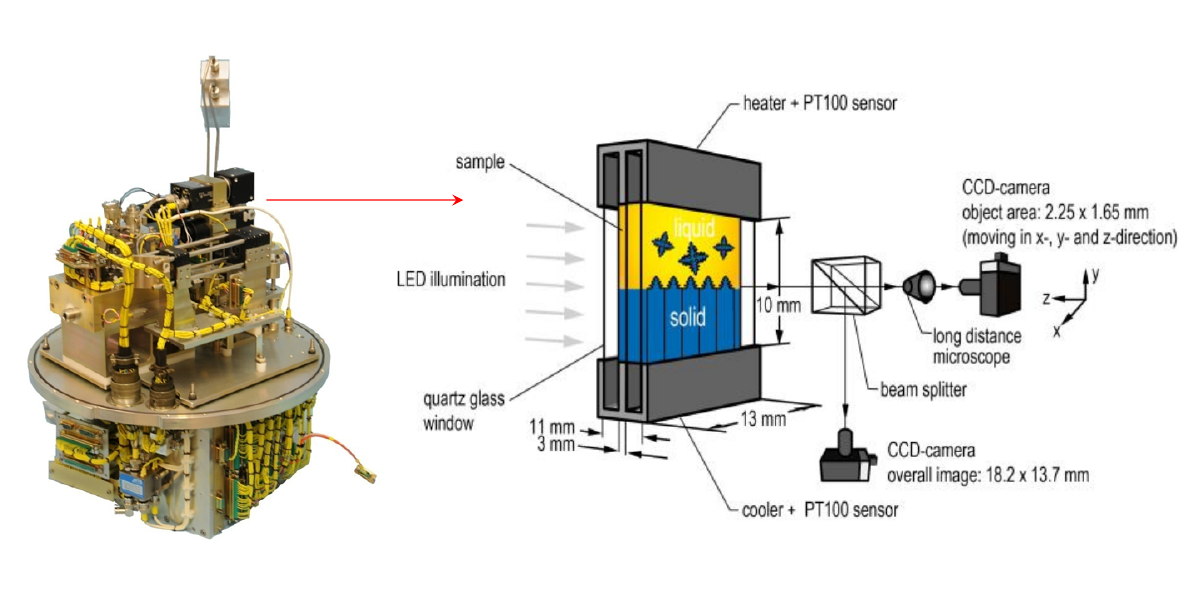}
  \caption{\label{texus_mod}Schematic representation of the TEM 06-23 TEXUS module\ref{pfeqxa}.}
\end{figure}

To investigate equiaxed growth of dendrites with $<$100$>$ crystallographic orientation the binary alloy
Neopentylglycol-20.0 wt.\% (D)Camphor (NPG-DC) was selected. The phase diagram and relevant alloy
properties are given in \cite{sturz17,zimmermann17}. The experiment MEDI-2 was carried out on board the sounding 
rocket TEXUS-55 mission in 2018, where convection of the melt and sedimentation of the dendrites is negligible. 
The experimental conditions were chosen such as to obtain an equiaxed dendritic structure with diffusive 
conditions for heat and mass transport during the microgravity period. The dendrites were observed in-situ 
with two different optical systems to analyze the global and the microscopic features of equiaxed 
solidification using the TEM 06-23 TEXUS module. 
As a result, equiaxed dendrites with six dendrite arms growing perpendicular to each other were detected,
which is characteristic for a $<$100$>$ crystallographic orientation. A schematic representation of 
the TEM 06-23 TEXUS module is shown in Figure \ref{texus_mod}. This hardware was provided by 
Airbus Defense \& Space. The temperature evolution in the solid or liquid alloy material was measured by
three Ni-CrNi thermocouples with diameter 0.25 mm, which were located in the centre vertical plane 
inside the experimental cell.

\begin{figure}[htbp]
  \centering
  \includegraphics[width=0.93\columnwidth]{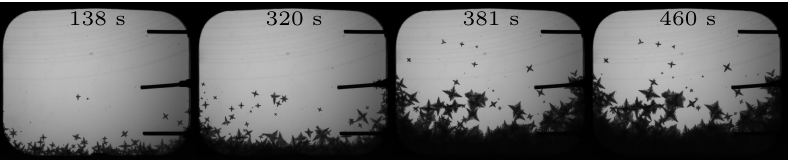}
  \caption{\label{texus_pic}Overview images of the experiment cell showing multiple equiaxed dendritic growth 
during the microgravity period. Also shown are the thermocouples TZ1 to TZ3 inserted from the right side.}
\end{figure}

Figure \ref{texus_pic} shows a series of 8-bit greyscale overview images taken with the CCD macro camera. 
The field of view is 13.6 mm in width and 10.9 mm in height, and the depth-of-field covers the full depth 
of the sample. The image acquisition rate is 10 frames per second with an optical resolution of 10.625 
$\mu$m/pixel. Also shown are the thermocouples inserted from the right side. 

At the beginning of the low-gravity period the NPG-DC alloy is completely liquid. Then, with the experimental
parameters chosen, a region with equiaxed dendrites, expands upwards in the experiment cell in a small thermal 
gradient. The \emph{in-situ} and real-time observation of this process allows for the investigation of the successive 
nucleation and growth of equiaxed dendrites, remaining at stationary positions in the microgravity environment 
(Fig. \ref{texus_pic}).

\section{Computer Vision Algorithm}

Our objective is to use deep learning neural network to identify and localize each dendrite on each experimental 
picture as shown in Fig \ref{texus_pic}. Several architectures already exists for this purpose. Some of the most 
applied are UNET \cite{unet}, Mask R-CNN \cite{maskrcnn} and Faster R-CNN \cite{fasterrcnn}. In this work, we will use 
Faster R-CNN because of its efficiency and for simple preparation of training data. 

\subsection{Faster R-CNN}



Faster R-CNN is one of the most popular object detection methods. It is part of the R-CNN series, developed by 
Girshick et al \cite{rcnn2014}, enhanced with Fast R-CNN \cite{fastrcnn} to finally obtain Faster R-CNN \cite{fasterrcnn}. 
We only describe the fundamentals of Faster R-CNN here.
The input image is passed through a convolutional neural network (CNN) to obtain a feature map of the objects present 
in the image. This part of Faster R-CNN's architecture is called the "backbone" network. This feature map is then used 
by a region proposal network (RPN) to generate region proposals (bounding boxes that contain the relevant image objects) using anchors 
(fixed-size reference boxes placed uniformly in the original image in order to detect the objects). These regions 
are then filtered by NMS (Non-Maximum Suppression), which is a method that allows to shift 
through the proposed regions and choose only those that are interesting. The feature map extracted by the CNN and 
the relevant object bounding boxes are used to generate a new feature map by pooling the regions of interest (RoI).
The grouped regions then pass through fully connected layers for object area coordinate prediction and output classes. 
This part of the Faster R-CNN architecture is called the header network.

\subsection{Training and testing}

The advantage of this method lies in the generic approach and can be extended to other microstructures if required. Furthermore, a delimitation of 
the dendrites both by a bounding box and a mask is automatically the outcome of the simulation. In the context of AI and R-CNN, the 
task is not to simulate the experimental NPG-DC dendrites as physically as possible, but to generate objects that look as similar as possible to the dendrites.

\section{Generation of training datasets}
 
\subsection{Phase field model}

Simulations were performed using the multi-phase field model with obstacle potentials implemented in the software MICRESS \cite{micress}. 
This model \cite{Eiken2006} is implemented using finite-difference correction method used to improve numerical accuracy \cite{Eiken2012}, 
and anti-trapping and mobility correction methods \cite{Carre2013}.

\subsection{Simulation parameters}

\begin{figure}[!htb]
   \begin{minipage}{0.18\textwidth}
     \centering
     \includegraphics[width=.9\linewidth]{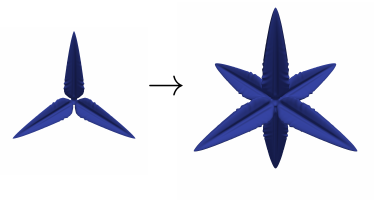}
     \caption{Simulation results at t=1s for 1/8 of a dendrite (isosurface for a phase fraction of 0.5), reconstruction of the whole dendrite by symmetry}\label{eqxreconstruct}
   \end{minipage}\hfill
   \begin{minipage}{0.78\textwidth}
     \centering
     \includegraphics[width=.9\linewidth]{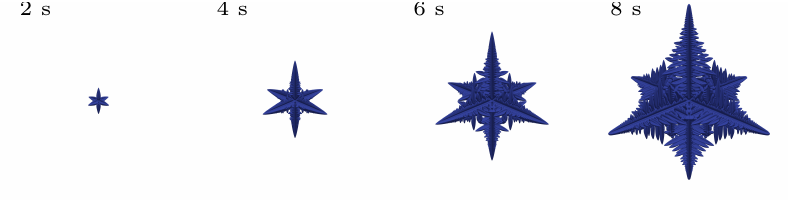}
     \caption{Simulated time evolution of reconstructed dendrite by symmetry}\label{eqxevol}
   \end{minipage}
\end{figure}

We chose a succinonitrile-1at\%acetone alloy for our simulations. Its
physical properties are given in Table \ref{tablethermophys}. Growth is initiated from a circular 
seed at the origin of a 3 dimensional cubic domain. The domain has  symmetric boundaries conditions 
on each face for phase field and alloy concentration. In this way we can simulate only one eight of a dendrite 
because of its internal symmetry and so reducing calculation time. After the calculation the dendrite is 
reconstructed by symmetry as we can see on the figure \ref{eqxreconstruct}. The domain length is 880 $\mu$m. 
The grid is dicretized on finite volume cubic element with a grid spacing of 1$\mu$m. The interface thickness 
is 4 $\mu$m. For equiaxed growth, isothermal conditions are used.

The results of one selected simulation are given on Fig. \ref{eqxevol} where time evolution of the dendrite morphology is shown. 
We have performed another simulation with a varying different interface properties. This lead us to a different morphological appearance 
of the equiaxed dendrite with less pronounced tertiary arms, as we can see on the figures  \ref{pfeqxa} and \ref{pfeqxb}, the dendrite 
arms are more smooth.

\begin{table}
\centering
\begin{tabular}{lc}
 Acetone concentration ($C_O$)  &  1at. \% \\
 Melting temp. of pure SCN ($T_f$) & 331.231 K \\
 Liquidus slope ($m_L$) & -2.1 K at.\%$^{-1}$ \\
 Partition coefficient ($k$) & 0.1 \\
 Diffusion coefficient in liquid ($D_l$) & 1.27$\cdot$10$^{-9} $  m$^2$s$^{-1}$\\
 Diffusion coefficient in solid ($D_s$) & 3$\cdot$10$^{-13} $ m$^2$s$^{-1}$ \\
 Gibbs-Thompson coefficient ($\Gamma$) & 6.525 $\cdot$10$^{-8}$ Km\\
 Interfacial energy ($\sigma_{sl}$) & 6.525 $\cdot$10$^{-2}$ Jm$^{-2}$
\end{tabular}
\caption{Thermophysical properties of the SCN-acetone alloy \cite{viardin2017}}
\label{tablethermophys}
\end{table}

\subsection{Generation of training and test dataset}

\begin{figure}[!htb]
   \begin{minipage}{0.48\textwidth}
     \centering
     \includegraphics[width=.9\linewidth]{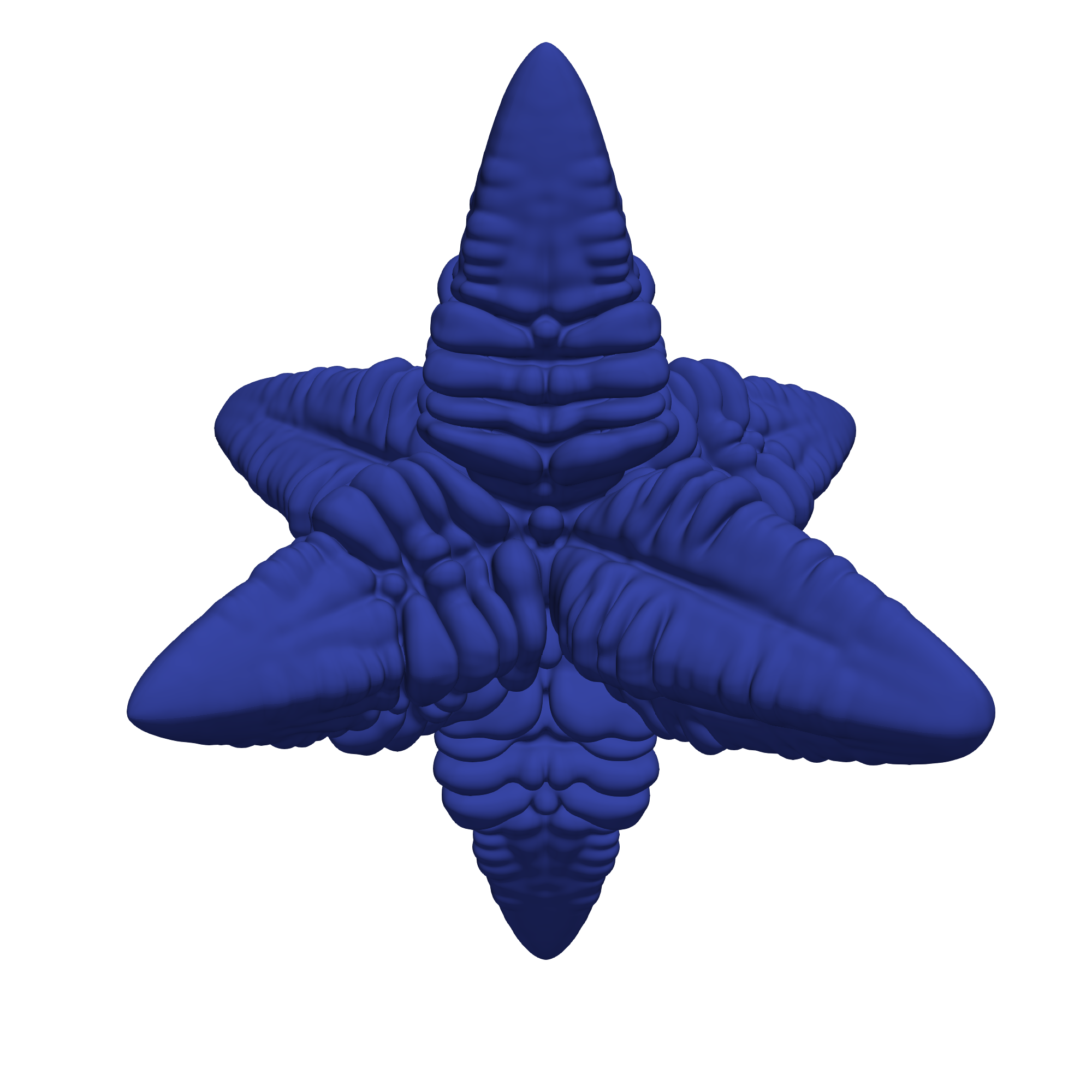}
     \caption{Artificial equiaxed dendrite generated with the phase field method. The interfacial energy anistropy ($\epsilon_{sl}$) is 0.7\% and the isothermal temperature is 328 K.}\label{pfeqxa}
   \end{minipage}\hfill
   \begin{minipage}{0.48\textwidth}
     \centering
     \includegraphics[width=.95\linewidth]{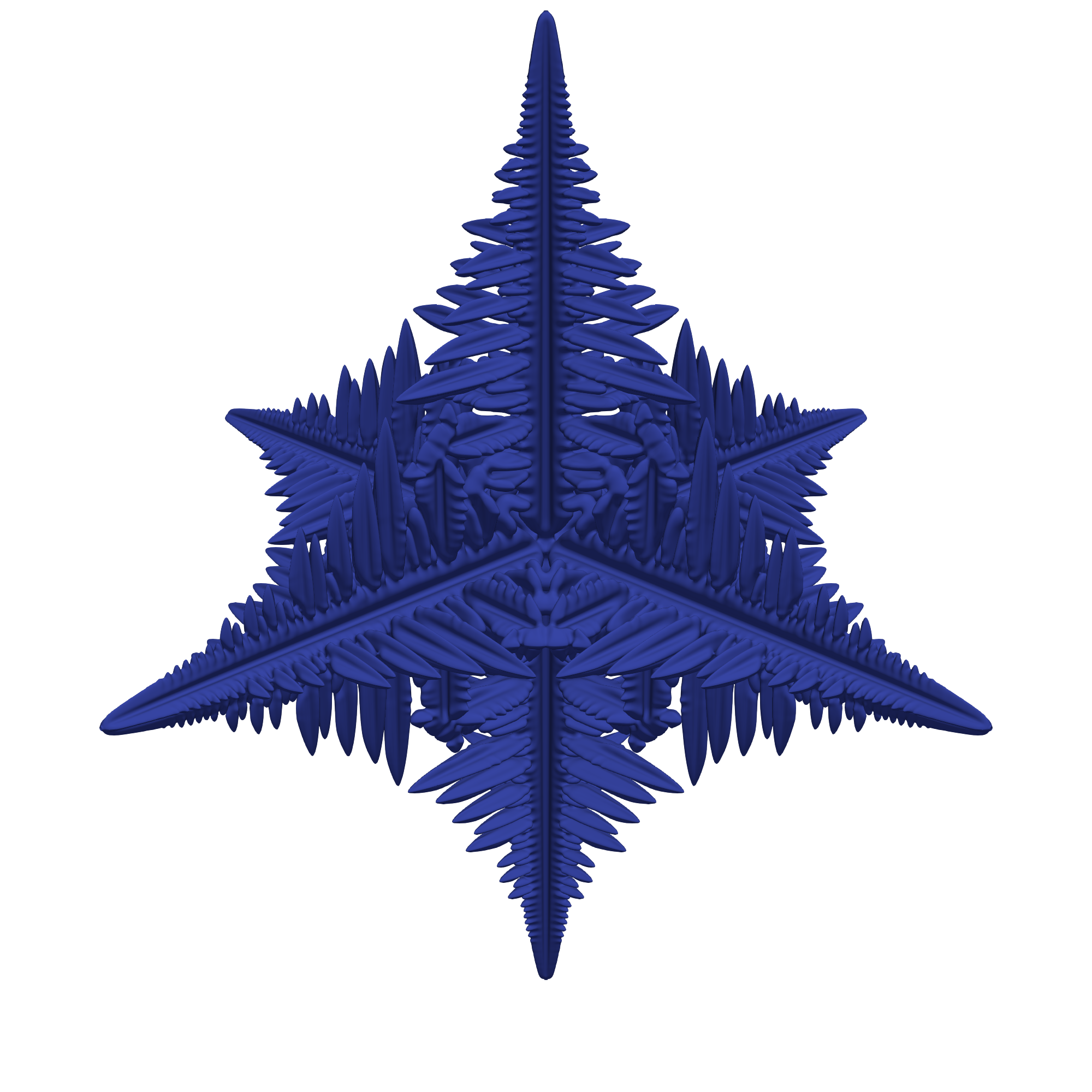}
     \caption{Artificial equiaxed dendrite generated with the phase field method. The interfacial energy anistropy ($\epsilon_{sl}$) is 0.07\% and the isothermal temperature is 328.67 K.}\label{pfeqxb}
   \end{minipage}
\end{figure}

The results of 3D simulations (isocontour for a phase fraction of 0.5 ) are presented on 
the figures \ref{pfeqxa} and  \ref{pfeqxb} for respectively anisotropy of interfacial energy 
value of $\epsilon$=0.105 and $\epsilon$=0.01, and isothermal temperature of T = 328 K 
(undercooling of 1.13 K) and T = 328.67 K (undercooling of 0.46 K).  
As we can see, the equiaxed dendrite on Fig. \ref{pfeqxb} have dendritic arms with smaller tip radii 
and needle like tertiary arms compared to \ref{pfeqxa}. The objective here is to have the two types 
of dendrites that we see in the experiments with differents kind of arm morphologies. 

\begin{figure}[htbp]
  \centering
  \includegraphics[width=0.93\columnwidth]{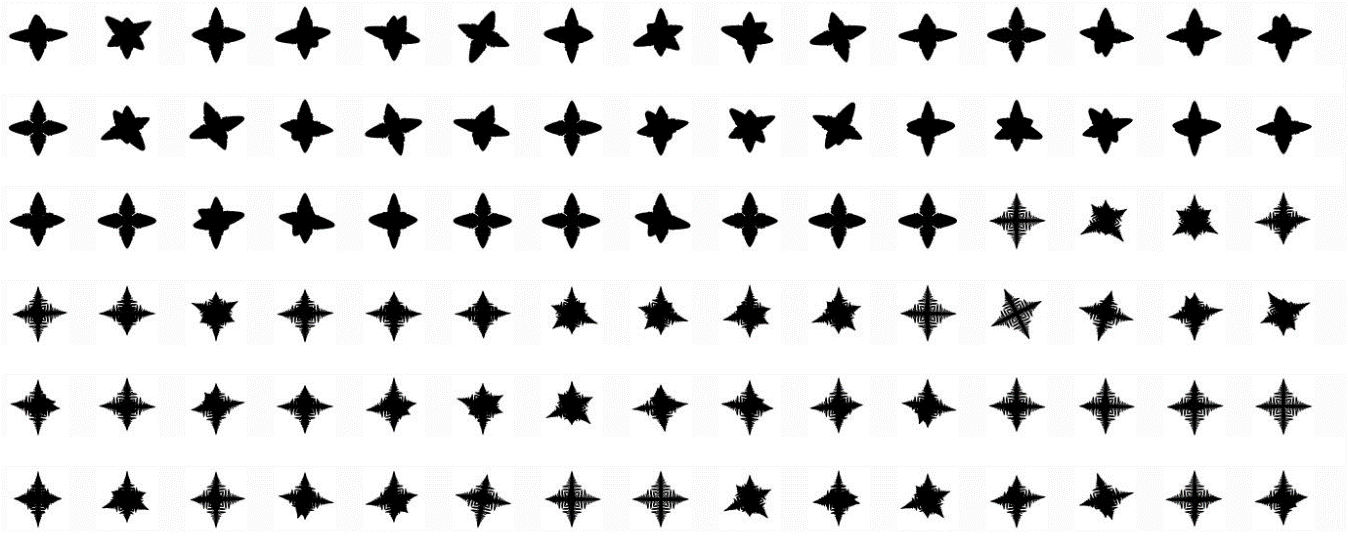}
  \caption{\label{pfeqproj}Different projections of the simulated 3D dendrites from Fig. \ref{pfeqxa}.}
\end{figure}

From these simulations results various 2D projections can be generated with different camera angle (using 
paraview scripting) as we can see on the figure \ref{pfeqproj}. 

\begin{figure}[!htb]
   \begin{minipage}{0.48\textwidth}
     \centering
     \includegraphics[width=.9\linewidth]{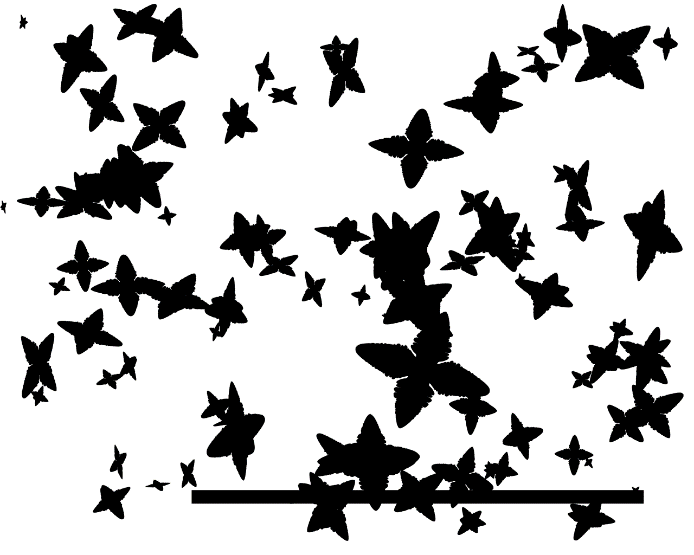}
     \caption{Example of a training image with augmented objects.\label{pftra}}
   \end{minipage}\hfill
   \begin{minipage}{0.48\textwidth}
     \centering
     \includegraphics[width=.95\linewidth]{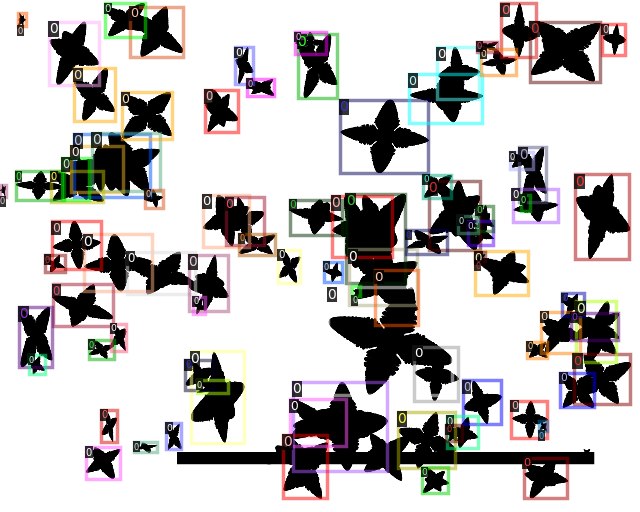}
     \caption{Example of a training image with augmented objects and automatically generated bounding boxes\label{pftrb}}
   \end{minipage}
\end{figure}

The projections are additionally stretched, overlapped, resized and rotated (augmentation) 
and anchored at random positions. A bounding box is generated and the class indexed ("1") 
for equiaxed dendrites. In addition, an artificial black bar was generated to depict 
the thermocouples. The detection of the thermocouples as separate objects has not yet 
been carried out. A result for a training image is shown in Fig. \ref{pftra}. The 
Fig. \ref{pftrb} is the same as Fig. \ref{pftra} but we have plotted the bounding boxes 
that are saved in the annotations files. 

\section{Faster R-CNN}

\subsection{Model implementation}

The Faster R-CNN architecture was implemented using the framework Detectron2, which is an open-source environment for AI. 
Detectron2 \cite{detectron2} is a library from Facebook$^\copyright$ that provides state-of-the-art algorithms for 
object detection and segmentation and often pre-trained general models. For our project 
we use a ResNet50 network with FPN (Feature Pyramid Network) as backbone.

\subsection{Training and testing parameters}

We have generated 80000 pictures, such as Fig. \ref{pftra} for training coming from phase field results. 
The dataset for testing were done using 10 pictures, like in Fig. \ref{texus_pic} coming from experiments with annotations.
The parameter for the Faster R-CNN architecture 
are listed in the Tab. \ref{tablefast}. The training was perfomed on a NVIDIA RTX 6000 GPU and last less than 
48 hours.

\begin{table}
\centering
\begin{tabular}{lc}
Image per batch  & 2\\
Base learning rate & 0.003\\
Warmup iterations & 100\\
Checkpoint Period & 10000\\
Maximum iterations & 130000\\
Steps & (30000,50000,100000)\\
Gamma & 0.5\\
Momentum & 0.9\\
ROI heads/Batch size per image & 128\\
Detections per Image & 150\\
Evaluation period & 1000\\
\end{tabular}
\caption{Parameter for the Faster R-CNN implemented with Detectron2 \cite{detectron2}}
\label{tablefast}
\end{table}


\subsection{Results of Faster R-CNN detection}

\begin{figure}[htbp]
  \centering
  \includegraphics[width=0.93\columnwidth]{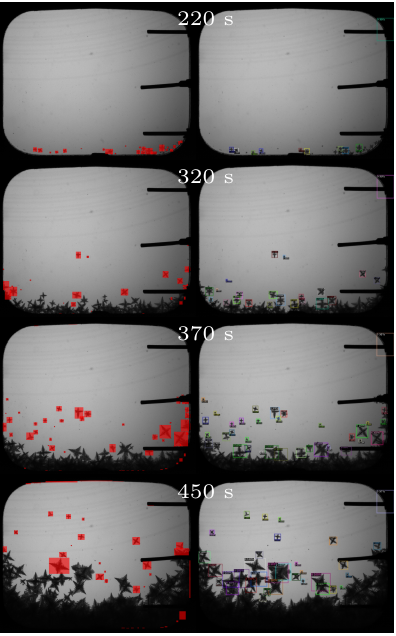}
  \caption{Comparison of classic detection, each on the left, and detection with faster R-CNN, each on the right, with selected image times:  220 s, 
  320 s, 370s, 450s.}
  \label{cmpall}
\end{figure}

On the figure \ref{cmpall}, we have plotted on the left column, the bounding box 
surrounding dendrites detected using traditional detection method (i.e with matlab) in red. 
On the right of the Fig. \ref{cmpall} we have plotted the bounding boxes surounding dendrites 
detected using Faster R-CNN. On each bounding box, the class probability is also given, it is in almost all of
the cases close to 100\%, this means that the probability that the detected object is a dendrite
is around 100\% which is true in our case. 

In general, the results are very promising. Nevertheless, small remarks have to be done  :
\begin{itemize}
\item{Very small dendrite at the beginning of experiments i.e 220 s are not well detected by Faster R-CNN}
\item{Faster R-CNN detects dendrites near the columnar front and overlapping dendrites much better than classical methods}
\item{There are artefacts with both methods: with the classic method, e.g. at the bottom right and with Faster R-CNN, a false 
positive object is detected in the upper right image area where the top thermocouple is located}
\item{When dendrites are strongly packed in the bottom of the experimental device, it is hard to detect them for the traditional method as well for
Faster R-CNN algorithm}
\end{itemize}

\begin{figure}[!htb]
   \begin{minipage}{0.48\textwidth}
     \centering
     \includegraphics[width=.9\linewidth]{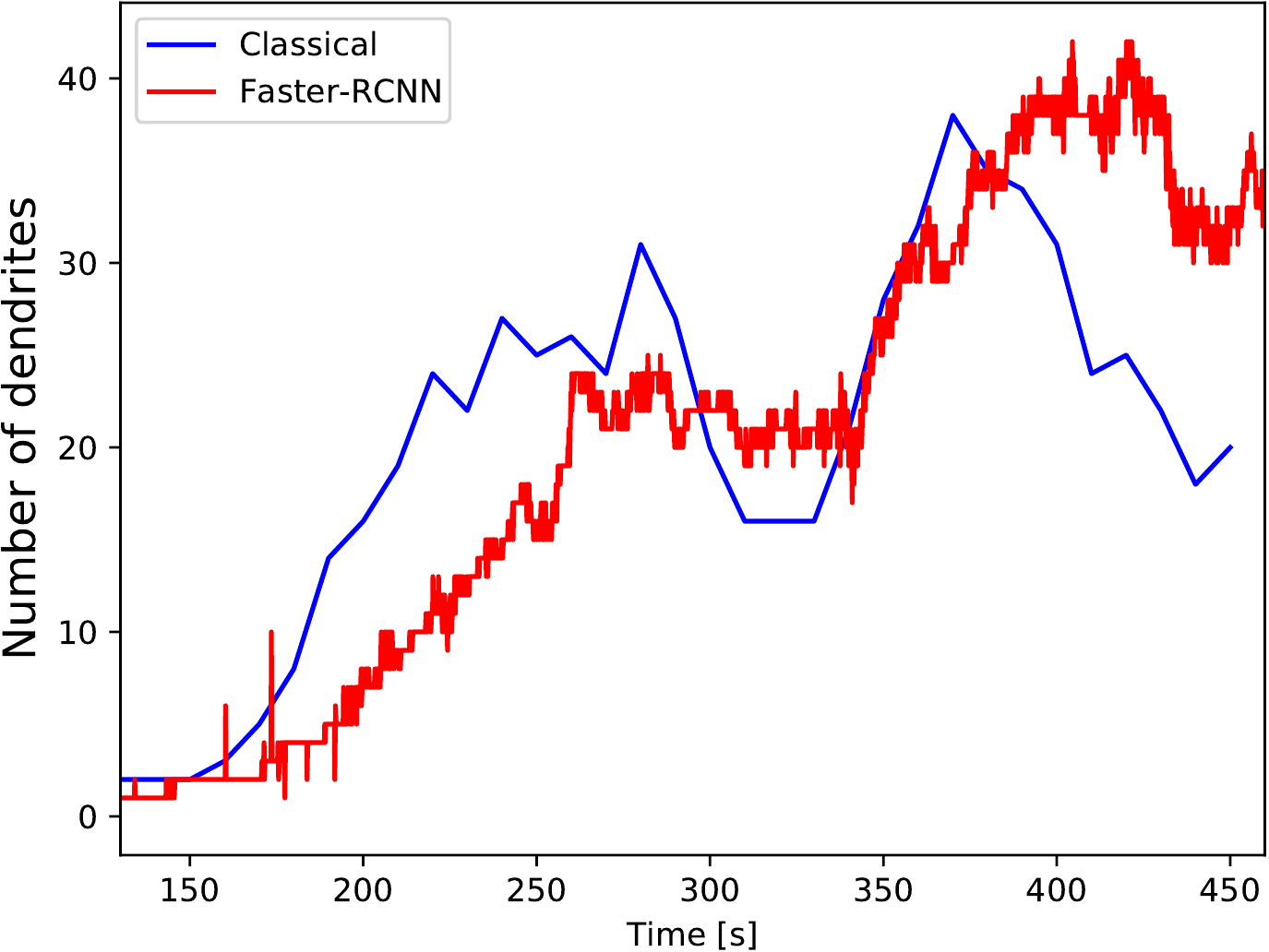}
     \caption{Number of detected dendrites from  of the experiment duration \label{rescmpa}}
   \end{minipage}\hfill
   \begin{minipage}{0.48\textwidth}
     \centering
     \includegraphics[width=.95\linewidth]{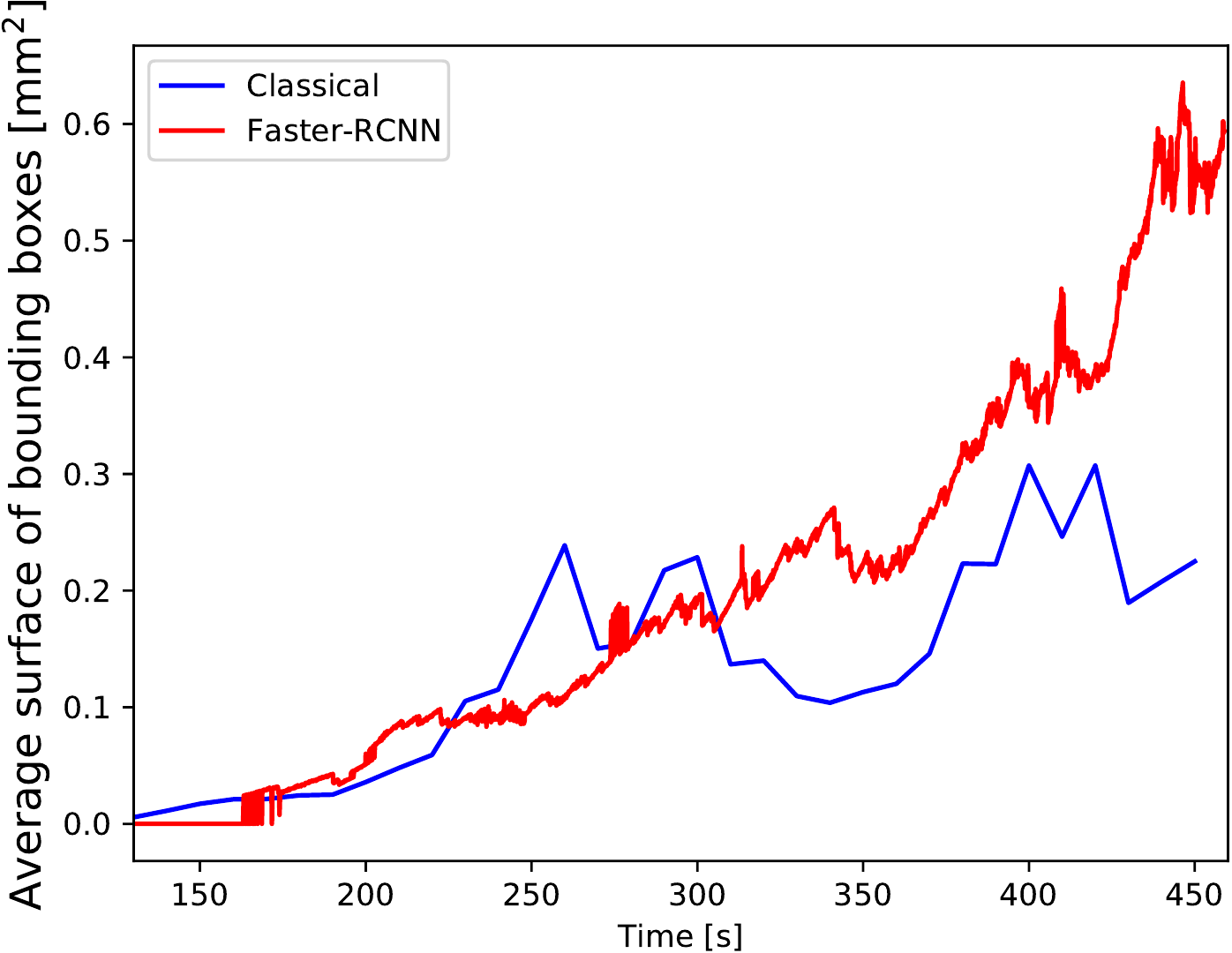}
     \caption{Determined box area of detected dendrites from as a function of the experiment duration \label{rescmpb}}
   \end{minipage}
\end{figure}

For a more detailed analysis, the evolution with time of the number of dendrites detected was evaluated for the classical method 
and Faster R-CNN and plotted in the figure \ref{rescmpa} as well for the average bounding box area on figure \ref{rescmpb}. 
The artifacts were manually removed concerning the classical method as well for Faster R-CNN. 
For the evolution of the number of dendrites with time, one can see at early stage i.e between 150 and 300 seconds the number of 
dendrites detected by the classical method is higher than Faster R-CNN because Faster R-CNN did not detect very small dendrites 
in our first model, while classical methods did. 
as classical method does. But at larger time i.e superior to 300 s, it is opposite. Faster R-CNN detects more dendrites than 
the classical method, at this stage Faster R-CNN detects overlapping dendrites better than the classical method and gives a realistic 
description of the number of dendrite in this experiment. The same observation can be made for the average bounding box area in figure 
\ref{rescmpb}, the better detection of packed/overlapping dendrites with Faster R-CNN method drive to higher average bounding box at latter 
stage of the experiments. 

\subsection{Conclusion}

In this work we have presented a computer vision detection algorithm based on the Faster R-CNN model trained using phase field model results
for experimental picture analysis. 
We have shown that the detection using Faster R-CNN provides better results than a conventional method using shape/edge 
detection. The Faster R-CNN was implemented using Detectron2 \cite{detectron2}.
The advantage is a limited number of phase field simulations or annotations of experimental images. Just with two phase field simulations we are able to generate a lot of data for training with realistically looking dendrites.  This method can be now applied to classify different solidification morphology like 
columnar dendrites or seaweed dendrites using training with phase field model results. 
    
\subsection{Acknowledgements}

This work was financially supported by DLR (Project KICK-G, contract number FKZ50WM2050).
The authors want to thank Dr. Markus Apel for fruitful discussion on phase field model.











\bibliographystyle{unsrt}
\bibliography{biblio}

\end{document}